\documentstyle[prl,aps,twocolumn,epsf]{revtex}

\draft
\newcommand{\r}{\vec r}

\begin{document}

\twocolumn[
\hsize\textwidth\columnwidth\hsize\csname@twocolumnfalse\endcsname

\title{{\large {\bf Persistent currents  on graphs }}}
\author{M. Pascaud and G.
Montambaux}
\address{Laboratoire de Physique des Solides, associ\'e au CNRS \\
Universit\'{e} Paris--Sud \\
91405 Orsay, France}
\date{December 30, 1998}
\maketitle

\begin{abstract}
{We develop a method to calculate the persistent currents and their spatial distribution
(and transport properties) on 
graphs made of quasi-1D diffusive wires.
They are directly related to the field derivatives 
of  the determinant of a 
matrix which describes the topology of the graph.
In certain limits, they are obtained by simple counting of the nodes and their 
connectivity. 
We relate the average current of a disordered graph {\it 
with 
interactions} and the {\it non-interacting} current of the same graph with clean 1D 
wires. A similar relation  
exists for orbital magnetism in general.
}
\end{abstract}

] \widetext
\narrowtext
The existence of  persistent currents in mesoscopic metallic rings is a thermodynamic signature of 
phase 
coherence\cite{exp}.
These currents  have been calculated using diagrammatic methods in which disorder and 
interactions are treated 
perturbatively\cite{theo,Schmid91,Ambegaokar90,Eckern91}, in a way very similar to the calculation 
of transport quantities 
like the weak-localization (WL)
correction, 
or the universal conductance fluctuations (UCF).
Like transport quantities\cite{Khmelnitskii84,Chakravarty86,Argaman96}, they have also been 
derived (after disorder averaging) using semi-classical 
calculations, in which they were expressed in terms of the classical and interference parts of the 
return probability for a 
diffusive particle\cite{Argaman93,Montambaux95,Gilles95,Ullmo97}. This formalism had made 
possible the calculation 
of WL corrections on any type of graph made of diffusive wires\cite{Doucot86}. Diffusion equation was solved on each 
link of the graph with 
current conservation on each node. For a network with $N$ nodes, the return probability could be 
related to the elements of 
a $N \times N$  "connectivity" matrix  $M$ and its inverse. This method has also been used recently 
to calculate the 
magnetization of such a network\cite{Pascaud97}, but it required rather lenghty 
calculations. 

In this letter, we show that the magnetization and the transport quantities
can be {\it directly} written in term of the determinant $\det M$ of the connectivity matrix. 
Besides being a very powerful 
method to calculate the 
above quantities, this result leads to  a straightforward harmonic expansion of these quantities for any network geometry. 
The efficiency of this 
method is shown for simple geometries of connected rings.
In addition, we are able to derive the
local distribution of the currents in the links of the network. Since the persistent current problem has still to be considered as unsolved, it is of interest to motivate new experiments in various geometries for which the magnetization and its distribution can be simply predicted and related to geometrical or topological parameters.

In the course of this work , we shall obtain a simple expression for the  spectral determinant 
of the diffusion 
equation, defined as:
\begin{equation}
{\cal S}_d (\gamma) = \prod_n b_n (\gamma + E_n)
\label{SD}
\end{equation}
where $E_n$ are the eigenvalues of the diffusion equation and $b_n$ are regularization 
factors\cite{Agam95}. Using the 
analogy between the diffusion and the Schr\"odinger equation, we will point out a very simple 
relation between the Hartree-
Fock (HF) average magnetization of a diffusive system and the 
grand canonical magnetization of the corresponding clean system. As a simple example, we relate 
the Aslamasov-Larkin 
contribution to the magnetization and the Landau susceptibility.

\medskip

All quantities of interest in this work can be related to the solution $P(\r,\r\,',\omega)$ of the diffusion equation in a magnetic 
field $\vec B= \vec \nabla \times 
\vec A(\r)$\cite{Cooperon}($\hbar=1$ throughout the paper):
\begin{equation}
\left \lbrack
- i \omega + \gamma - D(\nabla_{\vec r} -2ie \vec A)^2  \right \rbrack 
P(\r,\r\,',\omega)
=
\delta(\r-\r\,') 
\label{diffusion}
\end{equation}
$D$ is the diffusion constant. Unless specified, the magnetic field dependence is implicit. 
$\gamma= 1/\tau_\phi= D / L_\phi^2$  is the phase coherence rate. $L_\phi$ and $\tau_\phi$ are 
respectively the phase coherence 
length and time. 
 In the 
following, we will only 
need the space integrated return probability $P(t)=\int d^d\r P(\r,\r,t)$. It is simply written in 
terms of the 
eigenvalues $E_n$ of the diffusion equation, $P(t)=\sum_n e^{- (E_n+\gamma) t}=P_0(t) e^{- \gamma t}$. The time integral of 
$P(t)$, i.e. the Laplace 
transform of $P_0(t)$ can be straightforwardly written in terms of the spectral determinant 
(\ref{SD}). 
\begin{equation}
{\cal P} \equiv 
  \int_0^\infty dt  P(t) 
 = \sum_n \frac{1}{E_n + \gamma} 
= \frac{\partial}{\partial \gamma} \ln {\cal S}_d (\gamma)
\label{diff_spectral}
\end{equation}

Let us now recall how average magnetizations can be written in terms of $P(t)$. Here we restrict 
ourselves to $T=0K$.
The fluctuation of the magnetization 
$M _{typ} \equiv ( \langle M^2 \rangle - \langle M \rangle ^2)^{1/2}$ are given by\cite{Gilles95}:

\begin{equation}
M^2_{typ}  =
\frac{1}{2\pi^2}
\int_{0} ^{+\infty} dt
\frac{P''(t,B)-P''(t,0)}{t^3}
\end{equation}
where $P''(t,B)=\partial^2 P(t,B) / \partial B^2$. The main contribution to the average 
magnetization  is  due to electron-electron interactions\cite{Schmid91,Ambegaokar90}. Considering a screened interaction 
$U(\r-\r\,') = U \delta(\r-\r\,')$ and defining $\lambda_0 = U \rho_0$ where $\rho_0$ is the density of 
states (DoS) at the Fermi 
energy $\epsilon_F$, the Hartree-Fock (HF) contribution to the magnetization has be written 
as\cite{Montambaux95}:

\begin{equation}
\langle M _{ee} \rangle  = 
- \frac{\lambda_0}{\pi} 
\frac{\partial}{\partial B} 
\int_{0} ^{+\infty} dt \frac{P(t,B)}{t^2}
\end{equation}
Considering higher corrections in the Cooper channel leads to a ladder summation
\cite{Aslamasov74,Altshuler85b,Eckern91,Ullmo97}, so that $\lambda_0$ should be replaced by $\lambda(t)  = 
\lambda_0/(1+\lambda_0 
\ln(\epsilon_F t))$\cite{Sum}. We shall discuss later the contribution of this renormalization.

Using standard properties of Laplace transforms, the above time integrals can be written as  
integrals of the spectral 
determinant, so that the  magnetizations read:
\begin{eqnarray}
M^2_{typ}  & = &
 \frac{1}{2\pi^2}
\int_\gamma ^{+\infty} d\gamma_1
(\gamma - \gamma_1)\frac{\partial ^2}{\partial B ^2} \ln {\cal S}_d (\gamma_1) \left | _{0} ^{B} 
\right .
\label{Mtyp}
\\
\langle M _{ee} \rangle  & =  &
 \frac{\lambda_0}{\pi} 
\int_\gamma ^{+\infty} d\gamma_1
\frac{\partial}{\partial B}  \ln {\cal S}_d (\gamma_1)
\label{Mee}
\end{eqnarray}

In the case of a ring or a graph geometry, the integral converges at the upper limit. For the case 
of a magnetic field in a 
bulk system, this limit should be taken as $1/\tau_e$ where $\tau_e$ is the elastic time.
Finally, we  also recall that transport properties like WL or UCF can be also related to the 
spectral 
determinant\cite{Pascaud97}.

We now  wish to emphasize 
an interesting 
correspondence  between the HF magnetization of a phase coherent {\it interacting diffusive} 
system 
and the grand canonical magnetization $M_0$ of the corresponding {\it non-interacting clean} system. The 
latter can also be 
written  in term of a spectral determinant. The grand canonical magnetization $M_0$ is given quite generally by:
\begin{equation}
M_{0}  = 
-\frac{\partial \Omega}{\partial B}= - \frac{\partial }{\partial B}\int_{0} ^{\epsilon_F}
d\epsilon N(\epsilon)
\label{M1}
\end{equation}
where the integrated DoS is
\begin{equation} 
N(\epsilon)  =  
 -\frac{1}{\pi} \mbox{Im} 
\sum _{\epsilon_\mu} \ln (\epsilon_\mu - \epsilon_+)=-\frac{1}{\pi} \mbox{Im} 
 \ln {\cal S} (\epsilon_+)
\label{M2}
\end{equation}
 where $\epsilon_{+}=\epsilon+i0$, ${\cal S}(\epsilon)= \prod_{\epsilon_\mu} b_\mu(\epsilon_\mu - 
\epsilon )={\cal S}_d 
(\gamma=-
\epsilon)$.
where $\epsilon_\mu$ are the eigenvalues of the Schr\"odinger equation.

For a clean system these eigenvalues are the same as those of  the 
diffusion equation, with the
substitutions $D \rightarrow \hbar/(2 m)$ and $2e \rightarrow e$\cite{Boundaries}.

Comparing eqs.(\ref{M1},\ref{M2}) with eq.(\ref{Mee}), we can now formally relate $M_0$  and the HF magnetization $\langle M_{ee}\rangle$ of the same diffusive system:

\begin{equation}
M_{0} = - \mbox{lim}_{\lambda_0 \rightarrow 0} {1 \over \lambda_0} \mbox{Im}
[\langle M_{ee} \rangle (-\epsilon_F -i0) ]
\label{mapping}
\end{equation}
This limit corresponds to taking the first order contribution in $\lambda_0$. As a simple 
illustration, consider  the 
orbital magnetic susceptibility 
of an infinite disordered plane. For a disordered conductor, it is the Aslamasov-Larkin 
susceptibility $\chi_{AL}$ 
\cite{Aslamasov74}:

\begin{equation}
\chi_{AL} = \frac{4}{3} \frac{\hbar D}{\phi_0^2} \ln {\ln T_0 \tau_\phi \over \ln T_0 \tau_e}
\end{equation}
$T_0=\epsilon_F e^{1/\lambda_0}$ and $\phi_0=h/e$ is the flux quantum. After replacing $\gamma$  by $-\epsilon_F-i0$, taking the 
imaginary part of the 
logarithm and replacing $D$ and $2e$, we recover  the Landau 
susceptibility for the clean 
system:
$\chi_{0} = -e^2/(24\pi  m)$.
\medskip

We now calculate the spectral determinant for quasi-1D graphs. By solving the diffusion equation on 
each link, and then imposing 
Kirchoff type conditions on the nodes of the graph,  the problem is reduced to the solution of a 
system of $N$ linear 
equations relating the eigenvalues at the $N$ nodes. Let us introduce the $N \times N$ matrix 
$M$\cite{Sgraphs}:
\begin{equation}
M_{\alpha \alpha} = \sum_\beta \coth(\eta_{\alpha \beta})
\ \ , \ \
M_{\alpha \beta} = - \frac{e^{i \theta_{\alpha \beta}}}
{\sinh \eta_{\alpha \beta}}
\label{matrixM}
\end{equation}
The sum $\sum_\beta$ extends to all the nodes $\beta$ connected to the node $\alpha$; $l_{\alpha 
\beta}$ is the length of 
the link between $\alpha$ and $\beta$. $\eta_{\alpha \beta} = l_{\alpha \beta}/L_\phi$.
The off-diagonal coefficient $M_{\alpha \beta}$ is non zero only if there is a link connecting the 
nodes $\alpha$ and 
$\beta$. 
$\theta_{\alpha \beta} = (4 \pi /\phi_0)\int_\alpha ^\beta A.dl$ is the circulation of the vector potential 
between $\alpha$ and $\beta$. The authors of ref.\cite{Doucot86} derived a relation between ${\cal{P}}$ and the elements of the matrix $M$ and its inverse $T=M^{-1}$:

\begin{eqnarray}
2 \gamma {\cal P}&=&
(N-N_B) + 
\sum_{(\alpha \beta)} \eta_{\alpha \beta} F_{\alpha \beta}
\label{formule_doucot}
\\ 
F_{\alpha \beta}&=&
\coth \eta_{\alpha \beta}
-
{(T_{\alpha \alpha} + T_{\beta \beta})\over \sinh ^2 \eta_{\alpha \beta}}+ 2 
\mbox{Re}(e^{i\theta_{\alpha \beta}} T_{\beta 
\alpha})
\frac{\cosh\eta_{\alpha \beta}}
{\sinh ^2 \eta_{\alpha \beta}}
\nonumber
\end{eqnarray}
where $N_B$ is the number of links in the graph.
Using the  equality:
$\mbox{Tr} (M^{-1} \partial _\gamma M)
= \partial _\gamma \ln \det M$ and recognizing in each term of 
(\ref{formule_doucot}) the partial derivative with respect to $\gamma$,  we find that eq.(\ref{formule_doucot}) can  be rewritten as: $
{\cal P} =
\frac{\partial}{\partial \gamma}
\ln
{\cal S} _d $ where the spectral determinant ${\cal S} _d$ is given by:
\begin{equation}
{\cal S} _d = 
\left ( \frac{L_\phi}{L_0} \right ) ^{N_B -N} \ 
\prod _{(\alpha \beta)} \sinh \eta_{\alpha \beta} \ \det M
\label{det_spectral_Q1D}
\end{equation}
apart from a multiplicative factor independent of $\gamma$ (or $L_\phi$). $L_0$ is an arbitrary length.
We have thus transformed the spectral determinant which is an infinite product in a finite product 
related to $\det M$.

As an example, we  consider  a disordered ring of perimeter $L$, to which one arm of length $b$ is 
attached. The spectral 
determinant is equal to:

${\cal S}_d =
\sinh R y \ \sinh y \ + \ 
2 \cosh R y \ (\cosh y - \cos (4 \pi  \varphi))
$
where $\varphi=\phi/\phi_0$ is the ratio between  the flux $\phi$ threading the ring and the flux 
quantum. $y = (L/L_\phi)$ and  $R=b/L$. 
Thus the average magnetization is:

\begin{equation}
\langle M_{ee} \rangle = 
{\lambda_0 e D \over  \pi^2}
\int _{L\over L_\phi} ^{\infty} 
\frac{2 \sin 4\pi \varphi \ \ y d y}
{\tanh Ry \sinh y + 2 (\cosh y - \cos 4 \pi \varphi)}
\label{arm}
\end{equation}

If there is no arm ($R=0$), we retrieve the classical expression for the average magnetization of 
a disordered ring 
\cite{Oh91}. We notice that, in the limit $b \gg L_\phi$, the magnetization remains finite and is 
equal to $2/3$ of the 
single ring magnetization (for $L_\phi \lesssim L$, which corresponds to typical experimental 
values).

We want first to outline once more the connection between ballistic and disordered regimes. From 
eq.(\ref{arm}) and with 
the mapping (\ref{mapping}), $\gamma \rightarrow -E-i0$ and $L/L_\phi \rightarrow i k L$ where $k$ 
is the wave vector of 
the solutions of the Schr\"odinger equation, we immediately recovers the current in a one channel 
 ballistic 
ring\cite{anneau1bras}.

Let us come back to a  diffusive network made of connected rings. 
Experimentally,  the coherence length is of the order of the perimeter of one ring so that only a 
few harmonics of the flux 
dependence may be observed.  It is then useful to make a perturbative expansion. We split the 
matrix as $M=D-N$, where $D$ is a diagonal matrix: 
$D_{\alpha \alpha}=M_{\alpha \alpha} \approx z_\alpha $ to the lowest order in 
$L_\phi$ ($z_\alpha$ is the connectivity of the node $\alpha$);
$N_{\alpha \beta}=M_{\alpha \beta}
\approx 2 e^{-l_{\alpha \beta}/L_\phi} e^{i \theta_{\alpha \beta}}$. Expanding $\ln \det (I -D^{-1} N)=\mbox{Tr} 
[\ln (I -D^{-1} N)]$, we 
have:
\begin{equation}
\ln \det M
=\ln \det D -  
\sum_{n \geq 1} \frac{1}{n} \mbox{Tr} [ (D^{-1} N )^n]
\end{equation}
We call ``loop'' $l$, a set of $n$ nodes linked by $n$ wires  in a closed loop. The length $L_{l}$ 
of a loop $l$ is the sum 
of the lengths of the $n(l)$ links. The flux dependent part of $\ln {\cal S}$ can be expanded as:
\begin{equation}
\ln {\cal S} = -2 \sum_{\{ l \}} 
\frac{2}{z_1} \ldots \frac{2}{z_{n(l)}} e^{-L_{l}/L_\phi} \ \cos(4\pi \phi_{l} / \phi_0)
\label{dev_1erordre}
\end{equation}
$\phi_{l}$ is the flux enclosed by the loop $l$.

\begin{figure}[!ht]
\centerline{
\epsfxsize 8cm
\epsffile{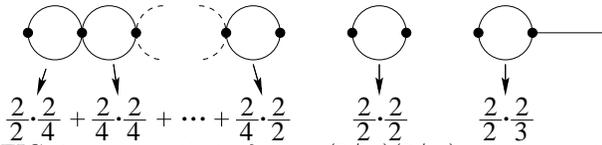}}
\caption{
connectivity factors $(2/z_1)(2/z_2)$ entering in the loop expansion (\ref{dev_1erordre}), for a series of 
identical connected rings, a 
single ring, and a ring with one arm.}
\label{Mee.chain}
\end{figure}

For example, we 
consider the cases shown on Fig.\ref{Mee.chain}. Reducing the above sum to elementary loops $l_0$ (with two 
nodes), so that $n(l_0)=2$, the first harmonics of the total magnetization, to the first order in 
$\lambda_0$ is:
\begin{equation} 
\langle M _{ee} \rangle = 2 G { \lambda_0 e D \over \pi^2} (L/L_\phi+1)  e^{-L/L_\phi}
\end{equation}
where $G=\sum_{\{l_0\}} 4/(z_1 z_2)$. $z_1$ and $z_2$ are the connectivity of the two nodes of 
each loop. The sum is 
made 
over the $m$ rings of the structure (see Fig.\ref{Mee.chain}). In particular, it is $G=(m+2)/4$ 
for an open necklace of $m$ 
rings and $G=m/4$ for a closed necklace. The same reduction factors were 
obtained for weak-localization corrections after lengthy calculations for $m=1,2,3,\infty$ in ref.\cite{Doucot86}. For the isolated 
ring, one recovers the 
known first harmonics\cite{Gilles95} and the above reduction factor $2/3$    for the ring with 
one arm.
 For an harmonic $p$ of the magnetization, corresponding to a winding number $p$ in the diffusion process, one should renormalize the interaction parameter 
because of the Cooper 
renormalization $\lambda=\lambda_0/(1+\lambda_0 \ln \epsilon_F/(\pi E_c/p^2))$\cite{Eckern91}. 

\begin{figure}[ht]
\centerline{
\epsfysize 5cm
\epsffile{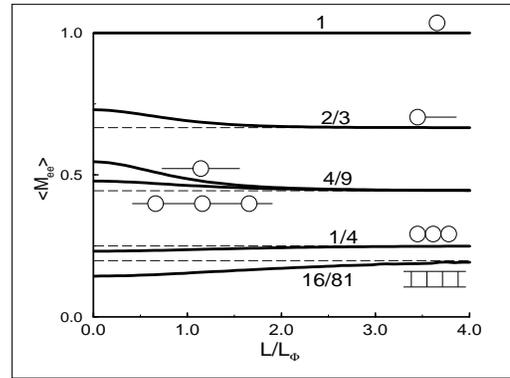}}
\caption{
Magnetization per ring for  networks of connected rings normalized to the single ring magnetization, calculated exactly (solid lines) 
and with the loop expansion 
(dashed lines).
The perimeter of all rings, and side-arm lengths are equal to $L$, The three bottom curves correspond to regular networks made of an infinite number of rings (only 
$3$ are represented). In 
these cases, the magnetization has been divided by the number of rings. The flux threading all 
rings is
$\phi=\phi_0/8$
}
\label{Mee.rapport}
\end{figure}

Fig.\ref{Mee.rapport}
displays a comparison between the magnetization of different networks of connected rings, evaluated numerically using eqs.(\ref{Mee},\ref{det_spectral_Q1D}).
The  
 perturbative expansions are in extremely good agreement with  exact results as soon as the 
coherence length is 
smaller than the perimeter of one ring (see dashed lines in Fig.\ref{Mee.rapport}).

Finally, we calculate the distribution of the local current on each link of the graph. On a link $(\alpha 
\beta)$, the average 
current is given by the derivative of the Hartree-Fock energy correction $E_{HF}$ to the vector 
potential $A(r), $where $r$ 
is any point belonging to the link $(\alpha \beta)$:

\begin{eqnarray}
&&\langle {\cal J}_{\alpha \beta}(r) \rangle
=- \frac{\delta E_{HF}
}{\delta A(r)} = \frac{\lambda}{\pi} \int_\gamma ^{+\infty} d \gamma_1
\frac{\delta \ln {\cal S}
}{\delta A(r)} \\
&&\frac{\delta \ln {\cal S}}{\delta A(r)} 
= \mbox{Tr} ( M^{-1} \frac{\delta}{\delta A(r)} M)
= \frac{16\pi}{\phi_0}   \mbox{Im} (M^{-1} _{\beta \alpha} 
M_{\alpha \beta}) 
\end{eqnarray}
Fluctuations of the  current corresponding to eq. (\ref{Mtyp}) can be obtained similarly\cite{these}.

\begin{figure}[!ht]
\centerline{
\epsfysize 5cm
\epsffile{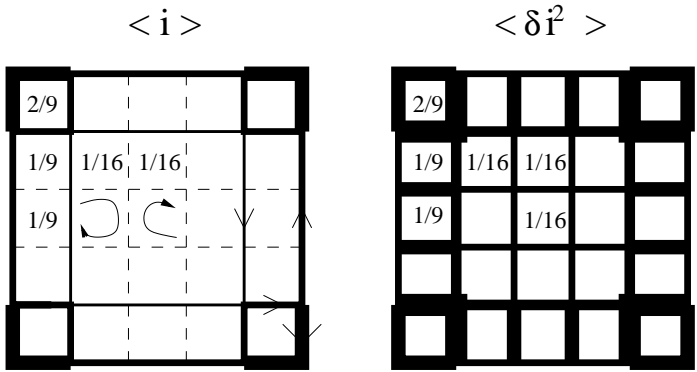}
}
\caption{Current distribution (left: average current; right: variance) for a square network, in the limit \protect{$L_\phi \leq L$}, i.e. when the flux dependence of the current is harmonic. The numbers show the amplitude of the average and typical magnetization per plaquette, in units of the magnetization of the single ring. It is maximum at the corner plaquettes. The thickness of each link is proportional to the amplitude of the current on this link, obtained by difference (resp. sum) of the average (resp. typical squared) magnetizations of the plaquettes neighboring the link. 
}
\label{distr_courant}
\end{figure}

In the limit $L_\phi \lesssim L$ considered above, the current distribution can also be derived 
quite simply. Indeed, in this approximation, the total magnetization can be written as a sum 
$\langle M_{ee} \rangle  \simeq  \sum_{k} \langle m_k \rangle$ where $\langle m_k \rangle$ is 
identified as the magnetization of 	a plaquette $k$ and  depends on the position of this 
plaquette in the array. It is given by the rules of eq. (\ref{dev_1erordre}) and is shown on 
Fig.\ref{distr_courant} for a regular square lattice. 
The average persistent current flowing in one link is the difference of the two plaquette currents 
neighboring it. The distribution of average 
current is sketched on Fig.\ref{distr_courant}.

The fluctuations can be described in the same way: namely it is a sum of terms which can be 
interpreted as fluctuations of the 
magnetization of one plaquette: thus the fluctuations of plaquettes are independent, and the fluctuations 
of current in one link is 
the sum of the fluctuations of its two nearby plaquette-current.
\medskip

In conclusion, we have developed a formalism which relates {\it directly} the persistent current, 
and the transport properties (although not detailed in this letter)  to the determinant of a 
matrix which describes the connectivity of the graph. From a loop expansion of this determinant, simple predictions for the magnetization and the spatial distribution of the  persistent current in any geometry can now be compared with forthcoming experiments on connected and disconnected rings. We have also found a correspondence between the phase coherent contribution to the orbital magnetism of a disordered 
interacting system and the orbital response of the corresponding clean non-interacting system.
\bigskip

We acknowledge useful discussions with E. Akkermans, A. Benoit,   E. Bogomolny, H. Bouchiat, D. Mailly and L. Saminadayar.


\begin{thebibliography}{99}
\bibitem{Cooperon} This solution is known as the Cooperon.
\bibitem{Boundaries} The boundary conditions may be different: for the isolated system trapped 
within infinite well, the 
Schr\"odinger equation  assumes Dirichlet boundary conditions while  the diffusion assumes the 
condition of zero current.
\end{thebibliography}
\end{document}